\documentclass[final,5p,times,twocolumn]{elsarticle}

\usepackage{hyperref}
\usepackage{siunitx}
\usepackage{graphicx}
\usepackage{subcaption}
\usepackage{multirow}

\journal{Journal of \LaTeX\ Templates}

\begin{document}

\begin{frontmatter}

\title{Innovative DC-coupled Resistive Silicon Detector for 4D tracking}

\author[UPO,INFN-TO]{R. Arcidiacono\corref{mycorr}}
\author[INFN-FI]{G. Bardelli}
\author[INFN-FI]{M. Bartolini}
\author[FBK,TIFPA]{M. Boscardin}
\author[INFN-TO]{N. Cartiglia}
\author[INFN-FI]{A. Cassese}
\author[FBK,TIFPA]{M. Centis Vignali}
\author[INFN-PG]{T. Croci} 
\author[INFN-TO]{\\ M. Ferrero}
\author[UniPG,INFN-PG]{A. Fondacci}
\author[FBK,TIFPA]{O. Hammad Ali}
\author[INFN-FI]{M. Lizzo}
\author[UPO,INFN-TO]{L. Menzio}
\author[INFN-PG]{A. Morozzi}
\author[CNR,INFN-PG]{F. Moscatelli}
\author[UniPG,INFN-PG]{D. Passeri}
\author[FBK,TIFPA]{G.Paternoster}
\author[INFN-FI]{G. Sguazzoni}
\author[INFN-TO]{F. Siviero}
\author[Unito,INFN-TO]{V. Sola}
\author[INFN-FI]{L. Viliani}

\address[UPO]{Universit\`a degli Studi del Piemonte Orientale, largo G. Donegani 2/3,28100 Novara, Italy}
\address[INFN-TO]{Istituto Nazionale di Fisica Nucleare, Sezione di Torino, via P. Giuria 1, 10125 Torino, Italy}
\address[INFN-FI]{Istituto Nazionale di Fisica Nucleare, Sezione di Firenze, Via Bruno Rossi 1, 50019 Sesto Fiorentino (FI), Italy}
\address[Unito]{Universit\`a degli Studi di Torino, via P. Giuria 1, 10125 Torino, Italy}
\address[CNR]{CNR-IOM, Sede secondaria di Perugia, via A. Pascoli 23c, 06123 Perugia, Italy}
\address[INFN-PG]{Istituto Nazionale di Fisica Nucleare, Sezione di Perugia, via A. Pascoli 23c, 06123 Perugia, Italy}
\address[FBK]{Fondazione Bruno Kessler, via Sommarive 18, 38123 Povo (TN), Italy}
\address[TIFPA]{TIFPA-INFN, via Sommarive 18, 38123 Povo (TN), Italy}
\address[UniPG]{Universit\`a degli Studi di Perugia, via G. Duranti 93, 06125 Perugia, Italy}

\cortext[mycorr]{Corresponding author \textit{Email address:}~roberta.arcidiacono@cern.ch}

\begin{abstract}
In the past 10 years, two design innovations, the introduction of low internal gain (LGAD) and of resistive read-out (RSD), have radically changed the performance of silicon detectors. The LGAD mechanism,  increasing the signal-to-noise ratio by about a factor of 20, leads to improved time resolution (typically 30 ps for a 50-$\mu$m thick sensor), while resistive read-out, sharing the collected charge among read-out electrodes, leads to excellent spatial resolution even using large pixels (about 15 $\mu$m for 450-$\mu$m pixel size). 

This contribution outlines the design strategy and presents the first performance results of the latest evolution of silicon sensors for 4D tracking, the DC-coupled Resistive Silicon Detector (DC-RSD). The DC-RSD is a thin LGAD with a DC-coupled resistive read-out. This design leads to signal containment within a predetermined number of electrodes using isolating trenches (TI technology). Several test structures and application-oriented devices have been implemented in the wafer layout. The sensors, produced at Fondazione Bruno Kessler (FBK) in the framework of the 4DSHARE project, have been characterized with a laser TCT system and recently tested at DESY with an electron beam.
The study of this first prototype production will provide us with immediate feedback on the soundness of the DC-RSD concepts.
\end{abstract}


\end{frontmatter}


\section{Introduction}
At the core of almost every high-energy physics (HEP) detector sits a silicon-based tracking system able to measure position and direction of the particles generated in the experiment.  Since their first appearance, silicon tracking detectors underwent a remarkably fast evolution, from a handful of electronics channels 40 years ago to the many millions of the current detectors, enabling many key discoveries. The requirements of future HEP experiments\cite{ecfa} point to spatial resolutions of just a few micrometers, timing resolutions of tens of picoseconds, low power consumption, and a reduced material budgets. 
 
With the introduction of Low Gain Avalanche Diode (LGAD) technology, silicon sensors have also gained a reputation for excellent timing resolution, opening the way to dedicated R\&D towards silicon 4D tracking sensors.
A promising solution able to meet the requirements for future 4D-trackers is represented by LGAD-based silicon sensors with a resistive read-out design. This solution has been initially implemented with an AC-coupled read-out, the so-called AC-LGAD sensors, or Resistive Silicon Detectors (RSD)\cite{RSDsecondProd}. The results obtained demonstrated that RSD sensors have the potential to reach the required accuracy in position and time. Performance studies of RSD matrices, with a limited number of pixels (3x3) and pitches in the range 100-500 micrometers, achieved a position resolution in the 5-15 $\mu$m range\cite{ARCIDIACONO2023168671,Menzio:2024khz}. 

These studies also showed three limitations of the AC-LGAD design: a non-uniform resolution across the sensor area, as it depends on the particle hit location with respect to the electrodes; signals are bipolar with rather long tails during the discharge; and possibly degraded performance in large-size devices or/and irradiated devices, as the leakage current of the whole sensor is collected uniquely at the periphery. 

\section{The DC-coupled Resistive Silicon Detectors}
The goal of the 4DSHARE project is to develop an improved version of the RSD sensor, suitable for the future 4D particle trackers, addressing the above mentioned shortcomings. 
The key points of the sensor design are: the implementation of built-in signal amplification (internal gain) and built-in signal sharing in a predetermined (and not too large) number of electrodes, to enhance the uniformity of response across the sensor surface. To this end, charge containment structures are introduced to define the pixel area and the signal sharing.  

\begin{figure}[htbp]
    \centering
    \includegraphics[width=0.99\linewidth]{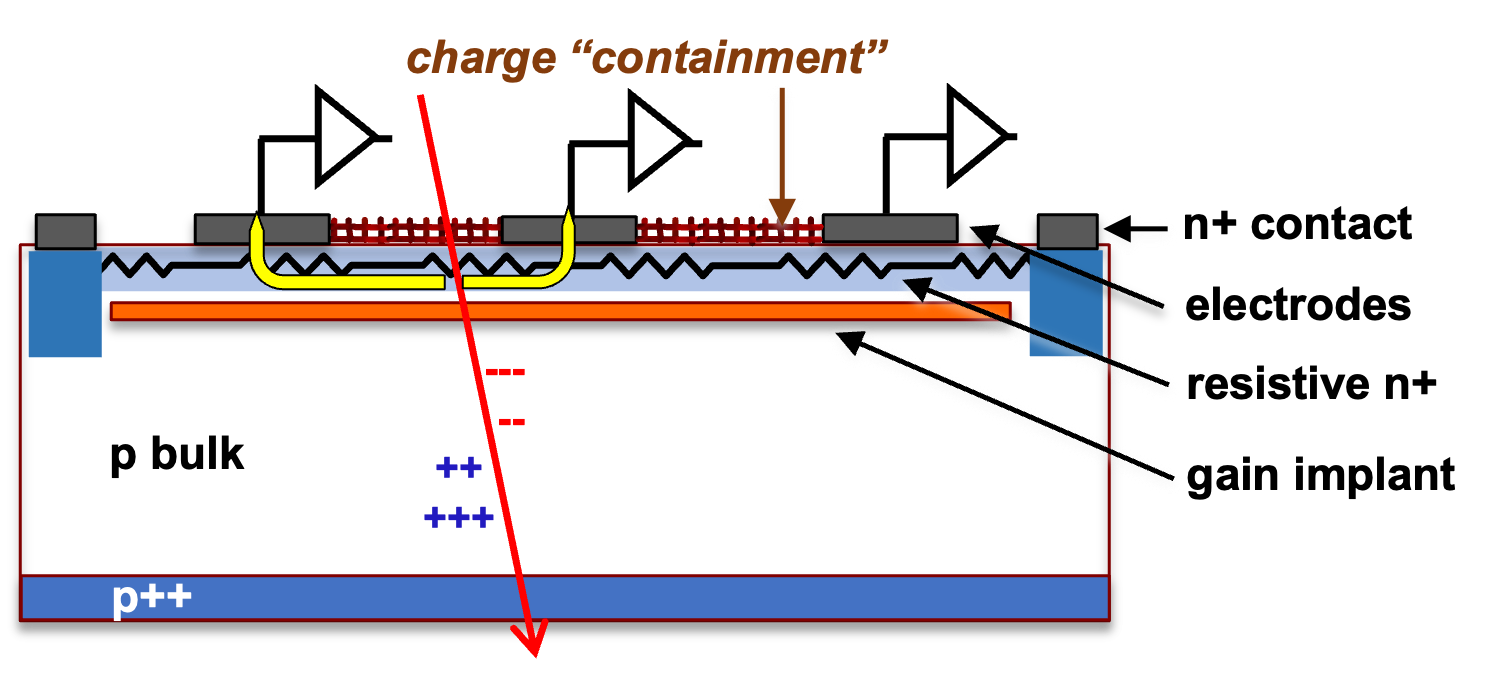}
    \caption{Sketch representing the crosscut of a DC-RSD sensor}
    \label{fig:DCRSD}
\end{figure}

The sensor is a thin LGAD with DC-coupled resistive read-out (DC-RSD), featuring 1-2 ns long unipolar signals and leakage current removal at each electrode. These characteristics are expected to ensure more uniform performance and a design that is scalable to large-area devices. Fig.\ref{fig:DCRSD}
shows a crosscut of a DC-RSD sensor: the read-out electrodes are implanted onto the resistive n$^+$ layer, and containment structures are added to confine the signals in a given area (pixel).

Extensive full 3D Technology CAD (TCAD) simulations have been performed to investigate the DC-RSD concept\cite{MOSCATELLI2024169380}, in terms of both electrical and transient behaviour. These simulations guarantees a very accurate evaluation of the sensor response to incoming ionizing particles. 
They have been instrumental for the evaluation of different technology options (e.g., the resistivity of the n$^+$ layer, contact materials) and geometrical layouts (shape and distance of the read-out electrodes). 

\section{The first DC-RSD prototype run}
The first DC-RSD prototype run, called {\it DC-RSD1}, comprises 15 p-type 6$"$ epitaxial wafers with an active thickness of 55 $\mu$m. The production features multiple technological options for the n$^+$ resistivity, gain implant dose and Si-Al DC-contact. The solution adopted for signal sharing containment within a pixel is the use of isolating trenches. The gain layer design is the shallow Boron implant implemented in standard FBK LGADs. The production has been completed at FBK in November 2024.

\begin{figure}[htbp]
    \centering
    \includegraphics[width=0.55\linewidth]{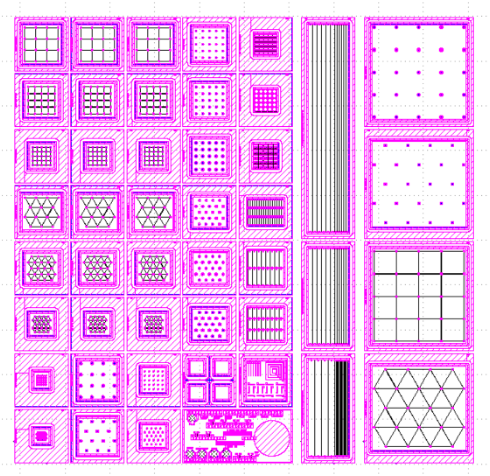}
    \caption{The reticle of the DC-RSD1 sensor production. Highlighted in dark-red the sensors used in this study}
    \label{fig:reticle}
\end{figure}

\begin{figure}[htbp]
    \centering
    \includegraphics[width=0.75\linewidth]{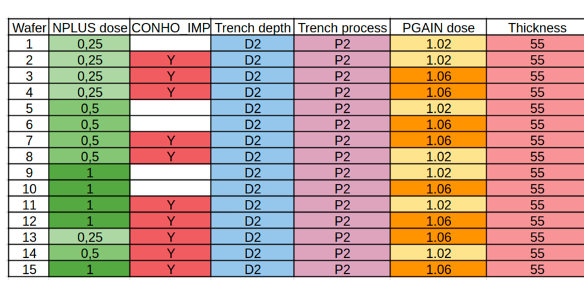}
    \caption{The split table of the DC-RSD1 sensor production}
    \label{fig:split}
\end{figure}

Several types of sensors have been designed, in multiple pitch options: devices with a square-shaped matrix of electrodes, with or without isolating trenches; devices with a triangle-shaped matrix of electrodes, with or without isolating trenches; strips of different length or electrode layout. Fig.~\ref{fig:reticle} and  ~\ref{fig:split} show the reticle and the split table of the DC-RSD1 production.


All wafer-level characterizations, performed by FBK, allowed to assess several parameters critical to the correct operation of the sensors, like the leakage current, breakdown voltage, the gain value, the resistivity of the n$^+$ layer and the DC-contact resistivity. Out of 15, seven wafers are fully functional, with the key parameters in the expected range. All the tested sensors types are functional. No mask layout mistake has been identified.

\section{Description of the test beam setup}

Three different trench-isolated DC-RSD types were tested at the DESY test beam facility, in the T22 experimental area, with a 5 GeV/c electron beam (December 2024): square-shaped pixels with 500-$\mu$m pitch, square-shaped pixels with 1000-$\mu$m pitch, triangle-shaped pixels with 500-$\mu$m side.

The T22 beamline is instrumented with a six Adenium planes telescope, three upstream and three downstream of the Detector Under Test (DUT). The telescope resolution, as used during this data-taking, has been estimated using the GBL calculator to be $\sigma_{x,y}$ = (8 $\pm$ 1.5) $\mu$m.

\begin{figure}[hbtp]
\begin{center}
  \includegraphics[width=0.8\linewidth]{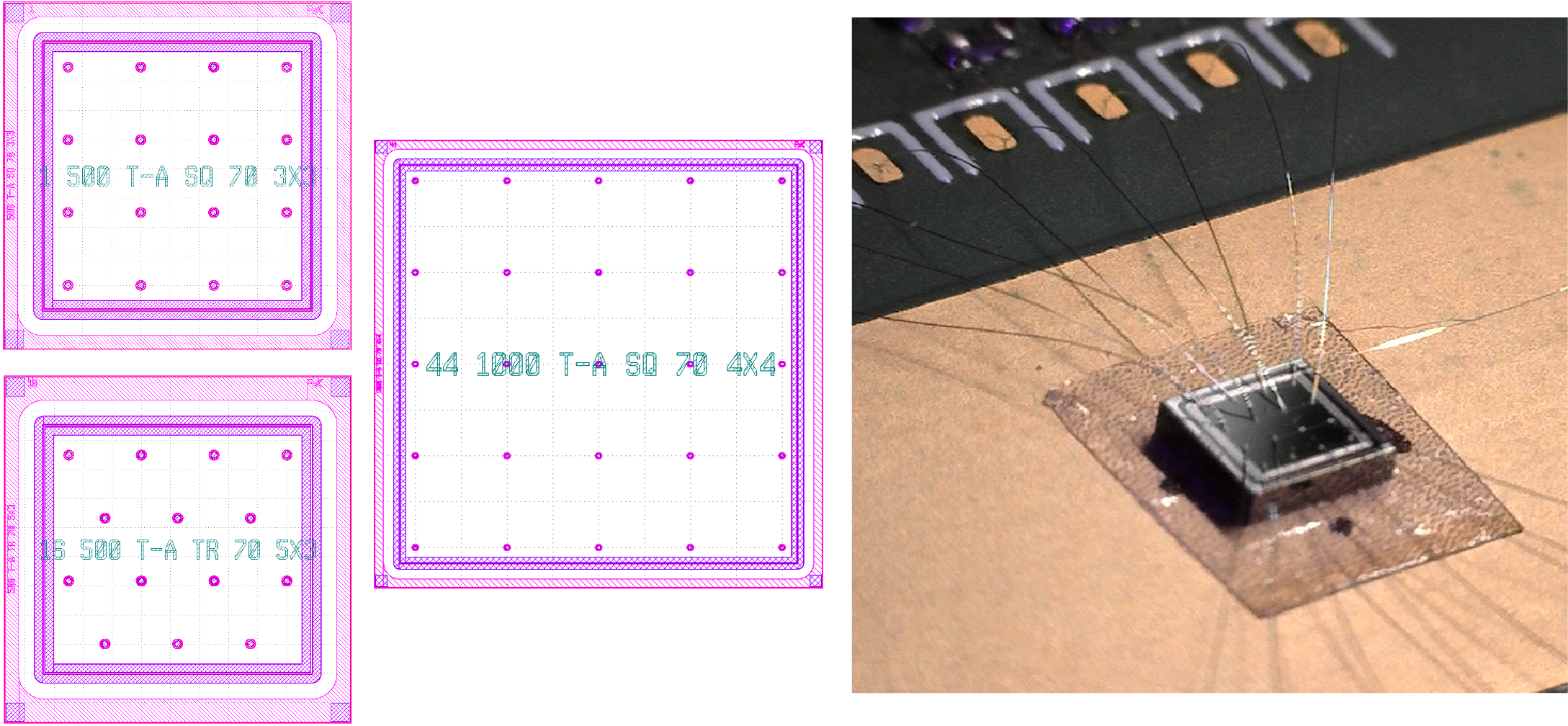}
\caption{DC-RSD sensor types tested at DESY (left). DC-RSD 3x3 pixels (500-$\mu$m pitch) matrix wirebonded to a FNAL board (right)}
\label{fig:sensorfoto}
\end{center}
\end{figure}

The experimental setup included, downstream the telescope hosting one DUT at a time, a large area thin LGAD, acting as trigger plane, and a Photonis MCP-PMT providing the reference timing to the particle track. The DC-RSD sensors to be tested were wire-bonded (see fig.~\ref{fig:sensorfoto}) to the {\it FNAL board}\cite{Apresyan_2020}, a 16-ch fast trans-impedance pre-amplifier analog board featuring a 2-stage amplifier chains based on the Mini-Circuits GALI-66+, well-suited for this type of sensors  thanks to its RC input circuit. The FNAL board was positioned inside a custom-made box fixed on the telescope PI stage. The temperature of the DUT was approximately 30-35$^\circ$C.

Two different acquisition modes have been used:
\begin{enumerate}
\item 15 DUT electrodes + MCP signal read out by a 16-ch CAEN  DT 5742 digitizer, for the position resolution measurements;
\item 7 DUT electrodes + MCP signal read out by an 8-ch LeCroy HDO9404 oscilloscope, for the time resolution measurements.
\end{enumerate}
The sensors under study have been selected from W3, which has the highest gain layer doping dose and n$^+$ resistivity. 
High-statistics runs have been taken at various reverse bias voltages to study the sensor performance under different operating conditions.

\section{DC-RSD signal properties}

\begin{figure}[htbp]
\begin{center}
  \includegraphics[width=0.9\linewidth]{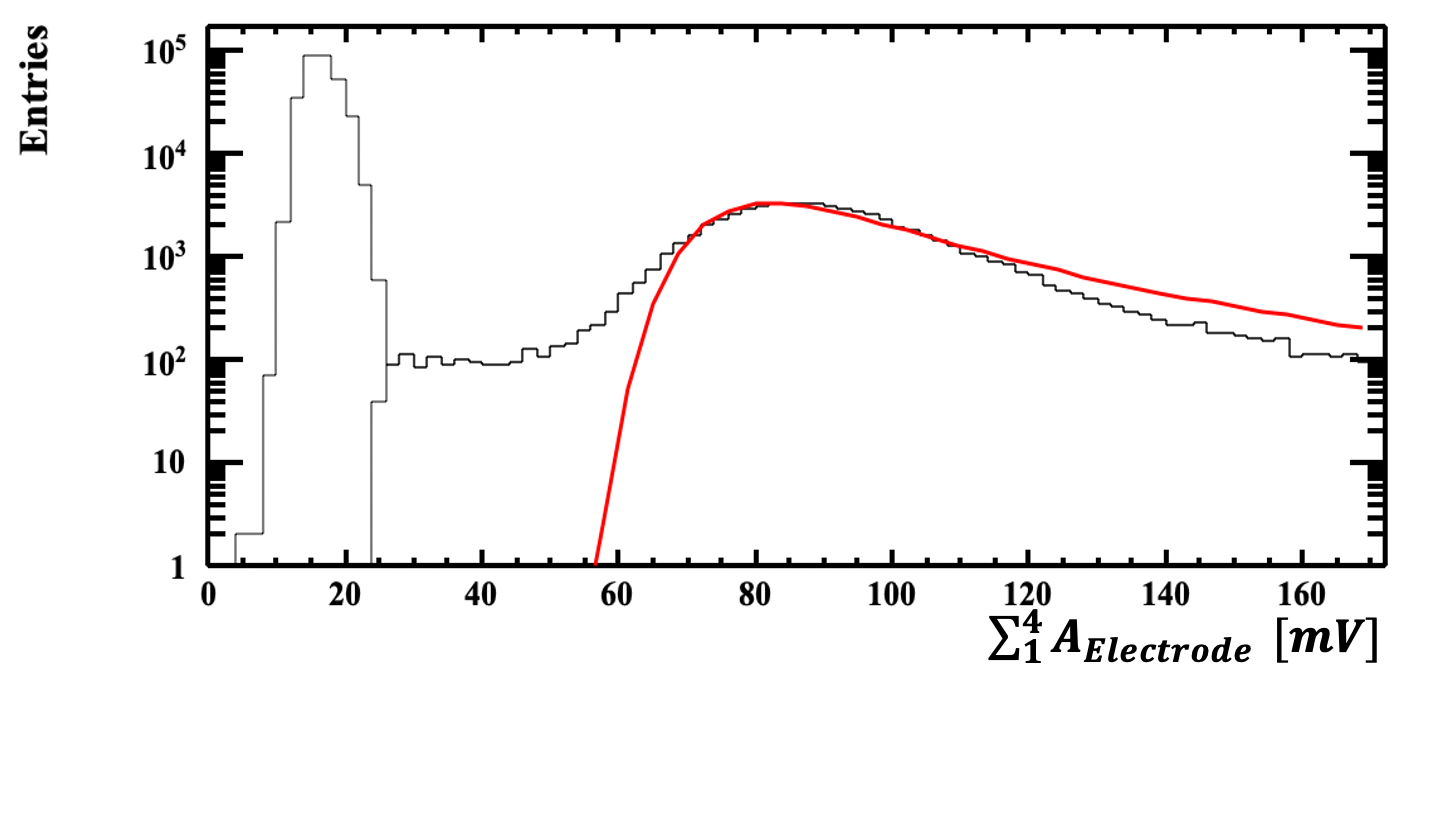}
\caption{Distribution of the total signal A$_{pixel}$ collected by a 500-$\mu$m pitch pixel biased at 240 V}
\label{fig:Amplitude}
\end{center}
\end{figure}

The signal A$_{pixel}$ collected by a DC-RSD {\it pixel} is estimated summing the amplitudes  A$_{i}$ (peak of the parabola fitted to the 6 highest samples) seen by the four (three) electrodes defining the squared (triangular) pixel shape. A$_{pixel}$ is shown in Fig.~\ref{fig:Amplitude} for the square-500-$\mu$m pitch sensor biased at 240 V: there is a clear separation between signal and noise. The event selection requires a A$_{pixel}$ $>$ 25 mV. Events with low amplitudes ( 25-55 mV range) are located close to isolating trenches. The data also show that the signal is seen only in the electrodes belonging to the pixel hit by the particle, indicating perfect charge containment.

\begin{figure}[htbp]
\begin{center}
  \includegraphics[width=1.0\linewidth]{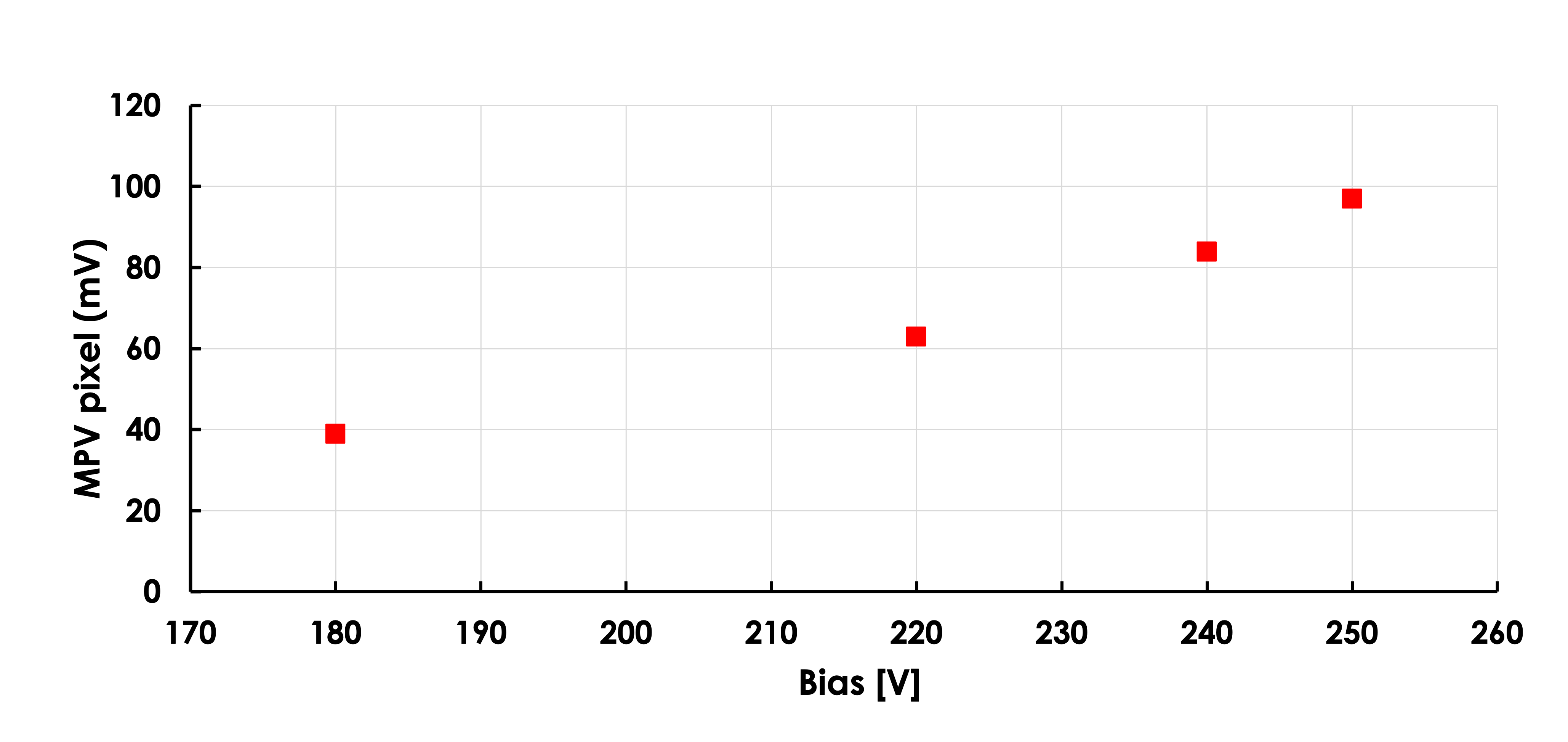}
\caption{MPV versus reverse bias voltage as measured for DC-RSD sensors (W3 - DC-RSD1) in the test beam setup at $\sim$30-35$^\circ$C.}
\label{MPVbias}
\end{center}
\end{figure}

In Fig.~\ref{MPVbias}, the most probable value (MPV) of the A$_{pixel}$ variable is shown as a function of the bias voltage applied. An MPV of 100 mV corresponds to a gain of about 35. The gain is estimated by measuring the trans-impedance of the FNAL board and the signal produced by a PIN diode read out with the same amplification chain. An uncertainty of 15\% is currently assigned to this computation. Dedicated measurements are ongoing in the laboratory to assess with higher accuracy the conversion factor MPV-gain in this setup. 

\section{Position and Time reconstruction: test beam results}

The hit position and time reconstruction procedures are the same as those used for the RSD sensors (template method), and have been described in detail in references~\cite{ARCIDIACONO2023168671,Menzio:2024khz}. In summary, look-up tables associating hit position with the signal-sharing fractions among the four (three) electrodes are produced, using an independent subset of test beam data taken with the square-shaped (triangle-shaped) sensor. 
For each event, the measured signal sharing fractions are compared with the look-up table to find the location that best reproduces the data.
The correlation between the positions measured by the tracker and by the DC-RSD sensor is very good over the whole read-out matrix.
The resolution is evaluated as the $\sigma$ of the Gaussian fit to the distribution of differences (x,y$_{DC-RSD}$ - x,y$_{tracker}$).

\begin{figure}[htbp]
\begin{center}
  \includegraphics[width=1.0\linewidth]{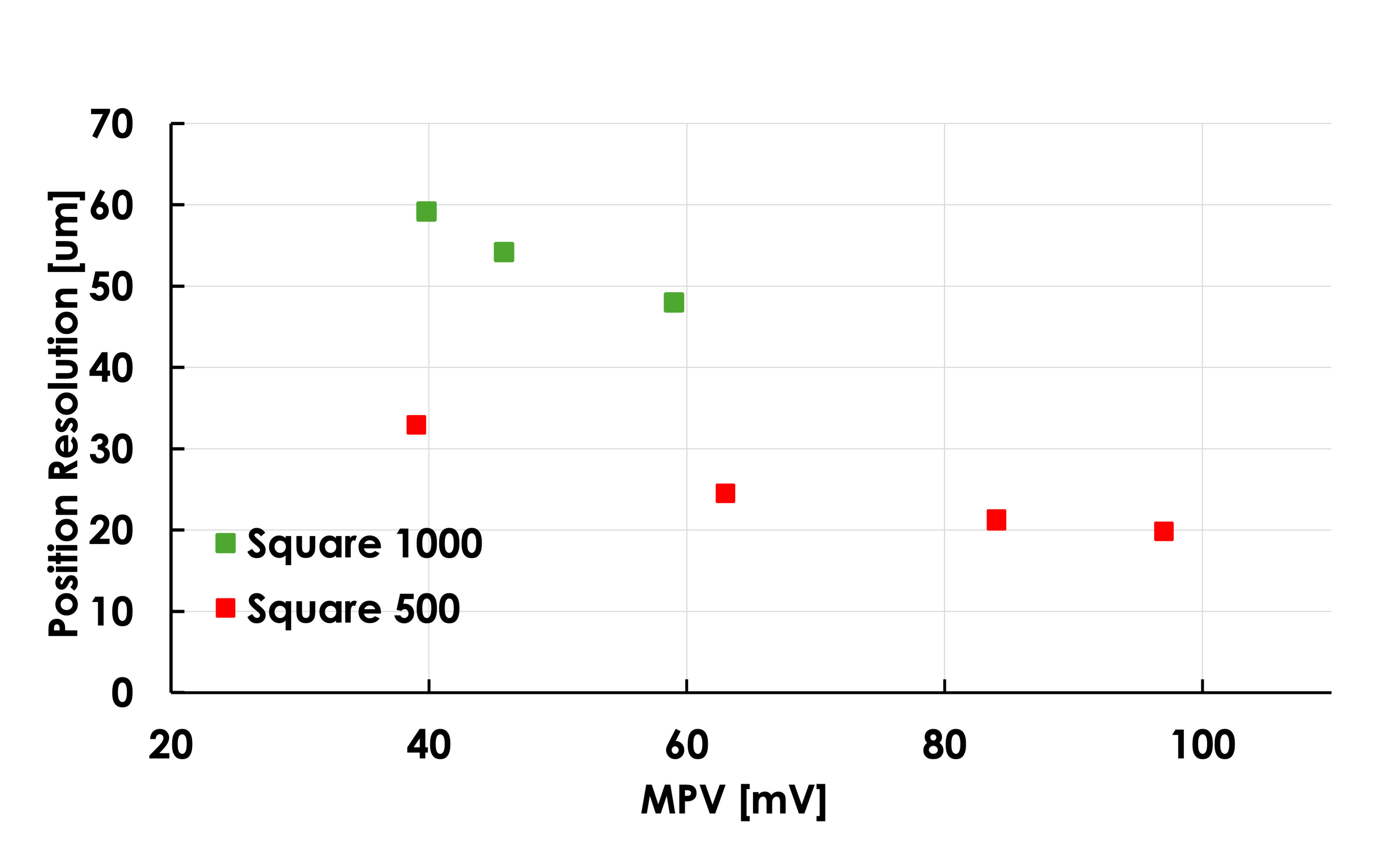}
\caption{Position resolution of DC-RSD sensors as a function of the MPV of the total signal amplitude measured by a pixel, for the square-shaped 500 and 1000-$\mu$m pitch sensors. It was not possible to bias the 1000-$\mu$m pitch sensor higher than 235 V}
\label{fig:spacevspitch}
\end{center}
\end{figure}

\begin{figure}[h]
\begin{center}
  \includegraphics[width=1.0\linewidth]{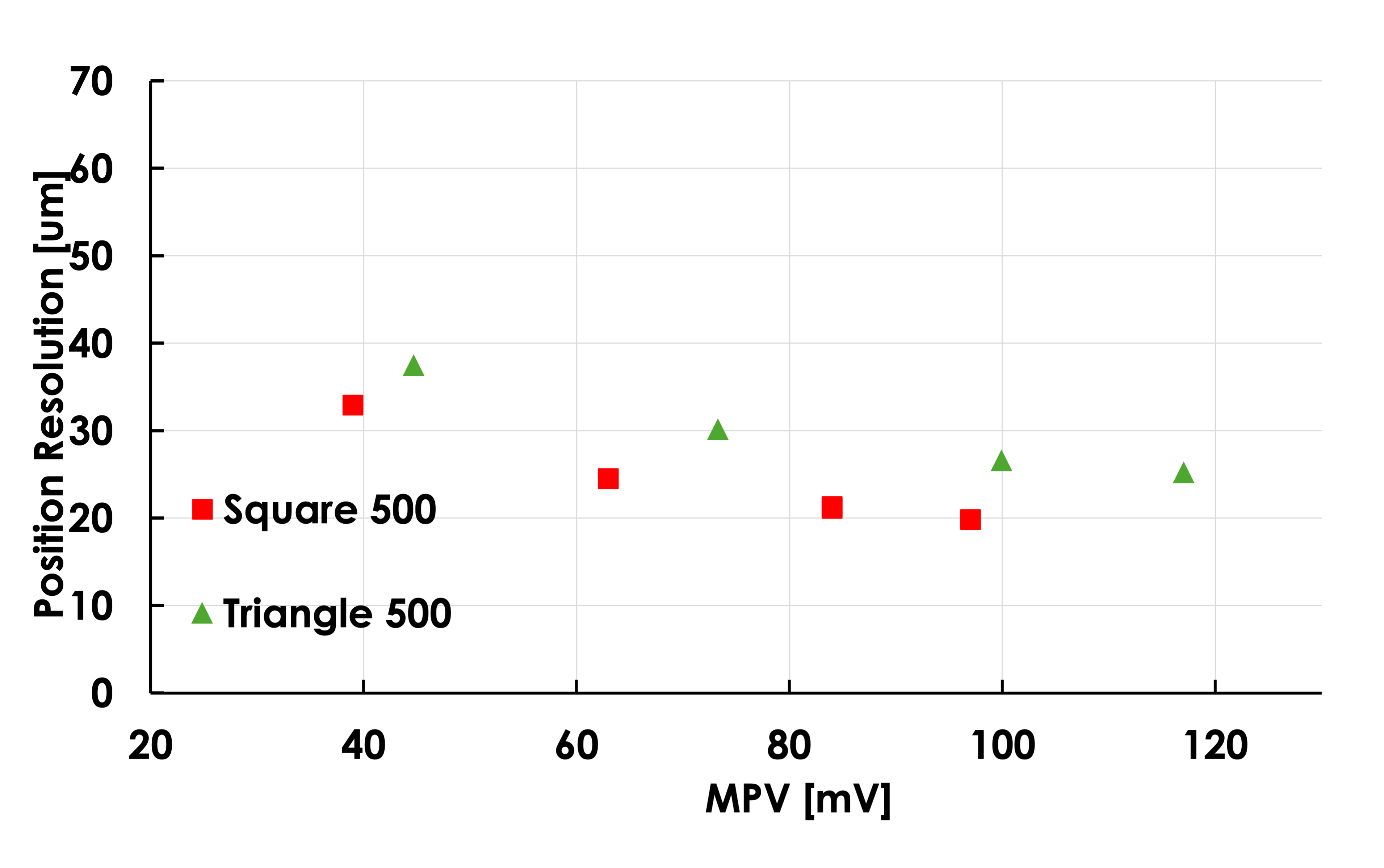}
\caption{Position resolution of DC-RSD sensors as a function of the MPV of the total signal amplitude measured by a pixel, for the square-shaped and triangle-shaped sensors. Square pixels perform better than triangular pixels.}
\label{fig:spacevsshape}
\end{center}
\end{figure}

The DUTs have been aligned w.r.t. the tracker; residual rotations have been estimated and corrected for; read-out chain amplification values have been inter-calibrated equalizing the MPV of the A$_{pixel}$ distributions.
Fig.~\ref{fig:spacevspitch} shows the position resolution at each bias voltage (as a function of the A$_{pixel}$ MPV) for the square-shaped 500 and 1000-$\mu$m pitch sensors, while fig.~\ref{fig:spacevsshape} compares the performance of square- with triangle-shaped pixels (500 $\mu$m pitch). In all plots, the telescope resolution has been subtracted in quadrature. All tested devices achieve a position resolutions better than 5\% of the pitch. 

For the hit time reconstruction, the time measured by each electrodes (using the constant fraction discriminator at 30\%) is corrected for signal propagation delay, which depends upon the hit position, and for a setup offset. The hit time t$_{DC-RSD}$ is computed combining the electrodes measurements. The time resolution is computed as the $\sigma$ of the Gaussian fit to the distribution of the differences (t$_{MCP}$ - t$_{DC-RSD}$). 

\begin{figure}[h]
\begin{center}
  \includegraphics[width=1.0\linewidth]{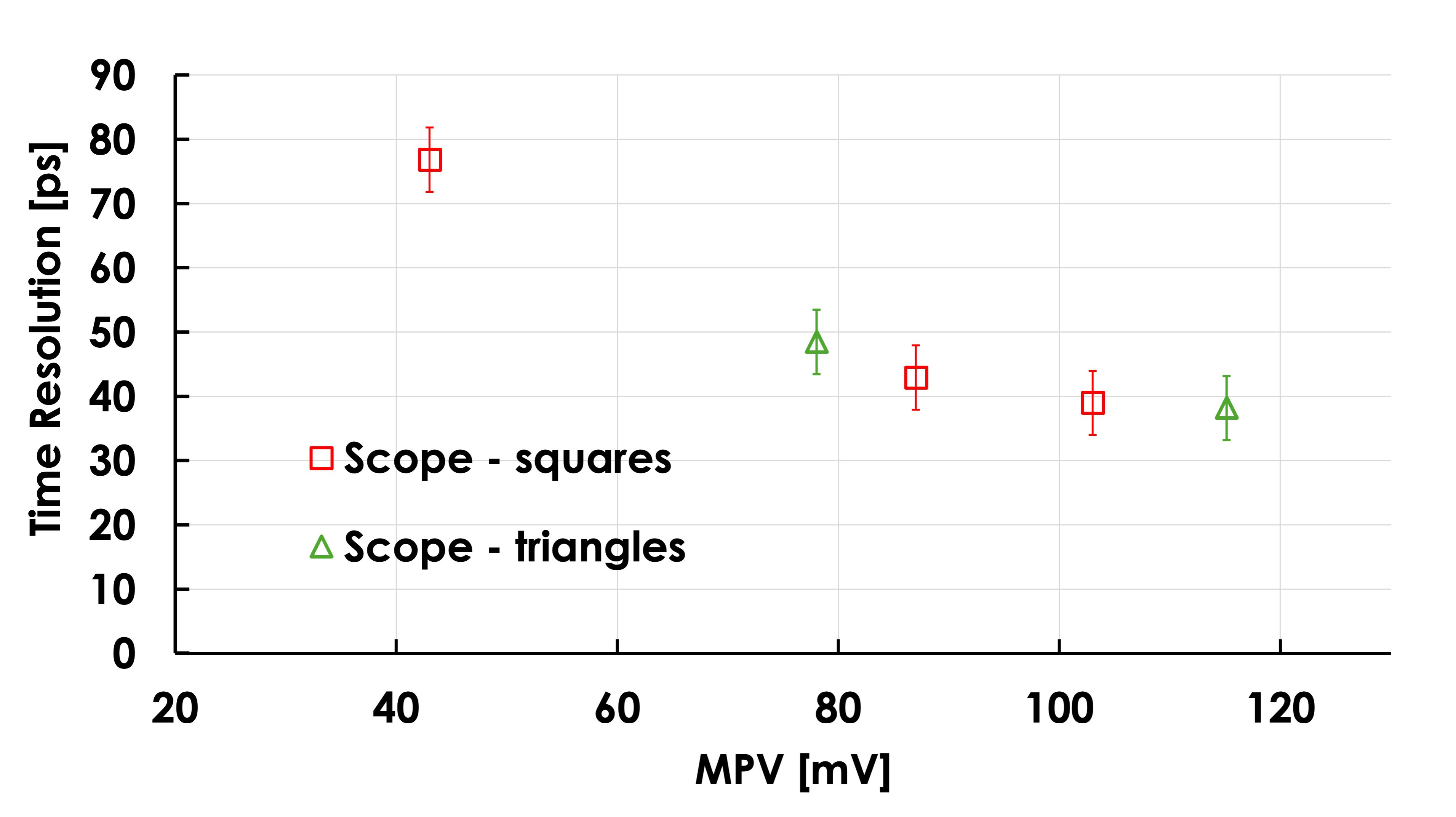}
\caption{Time resolution of DC-RSD sensors as a function of the MPV of the total signal amplitude measured by a pixel, for the square-shaped and triangle-shaped sensors. The MCP time resolution, estimated to be 5 ps, is subtracted in quadrature.}
\label{fig:time}
\end{center}
\end{figure}

Fig.~\ref{fig:time} shows the similar trend of the time resolution as a function of the A$_{pixel}$ MPV  for square- with triangle-shaped pixels (500 $\mu$m pitch). A time resolution $\sigma_{t} $ = 40 ps is achieved for MPV $>$ 90 mV). 

\section{Conclusions}
The first prototype run of DC-RSD sensors has been completed in late 2024.
The initial measurements done on-wafer at FBK provided important feedback on the various technological solutions implemented in the production.  
The main goals of the prototype run have been successfully achieved: manufacture functional trench-isolated resistive LGADs with DC-coupled electrodes collecting fast and unipolar signals, very similar to standard LGAD signals, propagating on the resistive n$^+$  layer. The electrodes have a sufficiently low resistance to efficiently collect the signals and the charge induced by an incoming ionizing particle is well contained by the trenches.

Preliminary measurements of space and time resolution are excellent. All tested devices (square-shaped pixels with 500, 1000-$\mu$m pitch and triangle-shaped pixels with 500-$\mu$m side) achieve a position resolutions better than 5\% of the pitch. As an example, a position resolution  $\sigma_{x,y}$ = 20 $\mu$m and a time resolution $\sigma_{t} $ = 40 ps, have been measured with the 500-$\mu$m pitch square-shaped pixel matrix operated at gain $\ge$ 30 (MPV $>$ 90 mV). 

A further improvement is expected with the ongoing optimization of the analysis procedure.
More data have been acquired using DC-RSDs with different pitch sizes and resistivity values. Complemented by the ongoing laboratory using a TCT laser setup, these data will lead to a thorough understanding of the signal formation and performance potential of these innovative 4D tracking sensors.

\section*{Acknowledgements}

We would like to acknowledge the support of the following Funding Agencies: INFN CSN5 through the 4DSHARE research project; European Union-Next Generation EU, Mission 4 component 2, CUP C53D23001510006 (project 2022KLK4LB); Compagnia San Paolo through the TRAPEZIO Grant; European Union's Horizon Europe Research and Innovation programme Grant Agreement No 101057511 (EURO-LABS). 
We acknowledge the fruitful discussions with the DRD3 collaboration, CERN. 


\bibliography{biblio}

\begin{thebibliography}{1}
\expandafter\ifx\csname url\endcsname\relax
  \def\url#1{\texttt{#1}}\fi
\expandafter\ifx\csname urlprefix\endcsname\relax\def\urlprefix{URL }\fi
\expandafter\ifx\csname href\endcsname\relax
  \def\href#1#2{#2} \def\path#1{#1}\fi

\bibitem{ecfa}
E.~D. R. R.~P. Group, {The 2021 ECFA detector research and development
  roadmap}, CERN-ESU-017 (2020).
\newblock \href {https://doi.org/10.17181/CERN.XDPL.W2EX}
  {\path{doi:10.17181/CERN.XDPL.W2EX}}.

\bibitem{RSDsecondProd}
M.~Mandurrino, et~al., {The second production of RSD\,(AC-LGAD) at FBK}, JINST
  17~(08) (2022) C08001.
\newblock \href {https://doi.org/10.1088/1748-0221/17/08/C08001}
  {\path{doi:10.1088/1748-0221/17/08/C08001}}.

\bibitem{ARCIDIACONO2023168671}
R.~Arcidiacono, et~al., {High-precision 4D tracking with large pixels using
  thin resistive silicon detectors}, Nucl. Instrum. Meth. A 1057 (2023) 168671.
\newblock \href {https://doi.org/10.1016/j.nima.2023.168671}
  {\path{doi:10.1016/j.nima.2023.168671}}.

\bibitem{Menzio:2024khz}
L.~Menzio, et~al., {First test beam measurement of the 4D resolution of an RSD
  pixel matrix connected to a FAST2 ASIC}, Nucl. Instrum. Meth. A 1065 (2024)
  169526.
\newblock \href {https://doi.org/10.1016/j.nima.2024.169526}
  {\path{doi:10.1016/j.nima.2024.169526}}.

\bibitem{MOSCATELLI2024169380}
F.~Moscatelli, et~al., {Measurements and TCAD simulations of innovative RSD and
  DC-RSD LGAD devices for future 4D tracking}, Nucl. Instrum. Meth. A 1064
  (2024) 169380.
\newblock \href {https://doi.org/10.1016/j.nima.2024.169380}
  {\path{doi:10.1016/j.nima.2024.169380}}.

\bibitem{Apresyan_2020}
A.~Apresyan, et~al., {Measurements of an AC-LGAD strip sensor with a 120 GeV
  proton beam}, Journal of Instrumentation 15~(09) (2020) P09038.
\newblock \href {https://doi.org/10.1088/1748-0221/15/09/P09038}
  {\path{doi:10.1088/1748-0221/15/09/P09038}}.

\end{thebibliography}

\end{document}